\newcommand{\definition}{\stackrel{\mathrm{d}}{\equiv}}
\documentclass[12pt]{iopart}
\usepackage{iopams}
  
\begin{document}

\title{Nonlinear completion of 
massive gravity of the Fierz-Pauli type}

\author{Shinji Hamamoto}

\address{Department of Physics, University of Toyama, Toyama 930-8555, Japan}
\ead{hamamoto@sci.u-toyama.ac.jp}
\begin{abstract}
A possible nonlinear completion of massive gravity of the Fierz-Pauli type 
is proposed. 
The theory describes a system consisting of a massive tensor field 
of the Fierz-Pauli type and an additional massive vector field. 
Massless limit as well as flat-spacetime limit can be taken smoothly. 
Constructing a nonlinear version of the physical-state condition 
which drives an extra scalar ghost from physical states 
is still unsettled. 
\end{abstract}

\submitto{JPhysA Special Issue-QTS5}
\maketitle

\section{Introduction}\label{sec:1}
In a series of papers \cite{SH1,SH2,SH3,SH4,SH5}, 
attempts to construct the theory of massive gravity 
with smooth massless limit were made. 

We studied infrared regularization
of linearized massive tensor fields in \cite{SH1,SH2,SH3,SH4}.  
Two model theories were considered: 
one is of the pure-tensor (PT) type, 
which describes an ordinary massive tensor field 
of five degrees of freedom; 
the other is of the additional-scalar-ghost (ASG) type, 
which contains a scalar ghost in addition to the pure tensor. 
The ASG model shows second-order massless singularities in 
two-point functions, whereas the PT model contains fourth-order 
singularities. 
It turns out that two procedures, the BRS one and the Nakanishi one, 
are effective in regularizing such singularities. 
The BRS procedure produces transparent structures to 
the resulting theories, 
as compared with the Nakanishi one. 
So we studied the former in detail.     
In order to drive away the second-order infrared singularities in 
the ASG model, we introduce an auxiliary vector-like field, 
and promote the original theory to the one 
that is invariant under the vector BRS transformation. 
On the other hand, to carry out infrared regularization of the 
fourth-order singularities in the PT model, 
we need to introduce an auxiliary scalar field  
in addition to the vector-like one, 
and make the resulting theory invariant under the scalar 
BRS transformation as well as the vector one. 

When we try to perform nonlinear completion,    
the ASG model is easier to deal with than the PT model. 
This is because only the vector BRS transformation 
is involved there. 
The nonlinear form of this transformation is simply 
the quantum version of the general coordinate transformation. 
The scalar BRS, on the other hand, has no classical counterpart.  
Constructing its proper nonlinear generalization 
is not an easy task. 
A possible nonlinear completion of the BRS model of 
the ASG-type massive tensor was proposed in \cite{SH5} 
(See also \cite{SH6}.). 
We also pointed out there that ghost condensation mechanism may 
work well for making innocuous the additional scalar ghost. 

The purpose of the present paper is to put into practice nonlinear 
completion of the infrared-regularized PT model. 
In order to avoid introducing scalar BRS, 
we ask for the help of the Nakanishi procedure. 
Then it is found straightforward to construct nonlinear Lagrangian 
for that BRS+Nakanishi model of the PT-type massive tensor. 

However, this is not the end of the story. 
In the present formulation, there occurs a new trouble 
concerning physical-state condition of the Nakanishi-type. 
Finding a nonlinear version of such condition requires 
further studies. 

In \sref{sec:2}, 
we review the case of Abelian vector field. 
This is to see how the BRS and the Nakanishi procedures work 
for regularizing massless singularities contained in the 
original massive theory. 
Stress is put on the fact that 
choosing massive gauge in the BRS procedure gives 
simple pole structure to two-point functions 
and makes it easy to investigate 
particle contents of physical states. 
In \sref{sec:3}, we treat linear theories of massive tensor field. 
Second-order massless singularities in the ASG model are regularized 
by the BRS procedure, 
whereas fourth-order singularities in the PT model are regularized by 
the use of both the BRS \textit{and} the Nakanishi procedures. 
Emphasis is laid also here on the usefulness of adopting massive gauge. 
\Sref{sec:4} treats nonlinear completion of the 
BRS+Nakanishi model of the PT-type massive tensor. 
After introducing nonlinear BRS transformation and basic BRS invariants, 
we propose possible nonlinear forms for the Lagrangian. 
Difficulties of finding a nonlinear version of the Nakanishi-type 
physical-state condition are also pointed out.  
Summary and discussion are given in \sref{sec:5}.  

\section{Massive vector}\label{sec:2}
\subsection{Massless vector}\label{sec:2.1}
Let us begin with massless vector.   
The Lagrangian is given by\footnote
{The flat spacetime metric used in the present paper is 
$\eta_{\mu\nu}=(-1, +1, +1, +1).$}
\begin{equation}
L_{0} = -\frac{1}{4}F_{\mu\nu}F^{\mu\nu} 
+ L^{\alpha}_{\mathrm{GF+FP}} , 
\label{eq:1}
\end{equation}
where $L^{\alpha}_{\mathrm{GF+FP}}$ 
is the gauge-fixing and Faddeev-Popov (GF+FP) Lagrangian  
\begin{eqnarray}
L^{\alpha}_{\mathrm{GF+FP}} & = 
b\left(\partial_{\mu}A^{\mu}
+ \frac{\alpha}{2}b\right) 
+\rmi\bar{c}\,\square\,c \nonumber\\
& = -\rmi\delta\left[\bar{c}
\left(\partial_{\mu}A^{\mu}
+ \frac{\alpha}{2}b\right)\right] 
\label{eq:2}
\end{eqnarray}
with the gauge parameter $\alpha$. 
The theory is invariant under the BRS transformation
\begin{equation}
\delta A_{\mu} = \partial_{\mu}c , \ \ \ \ 
\delta\bar{c} = \rmi b . 
\label{eq:3}
\end{equation}
For $\alpha=1$, two-point functions take simple forms:\footnote
{Here and hereafter the spacetime coordinates are omitted 
in the field variables as well as in the $\delta$-functions.} 
\begin{equation}
\langle A^{\mu}A^{\nu}\rangle =  
\frac{\eta^{\mu\nu}}{\Box}\delta ,\ \ \ \  
\langle A^{\mu}b\rangle = 
\frac{\partial^{\mu}}{\Box}\delta ,\ \ \ \ 
\langle bb\rangle = 0 ,\ \ \ \ 
\langle c\bar{c}\rangle = -\rmi\frac{1}{\Box}\delta . 
\label{eq:4}
\end{equation}
Physical states are defined by the use of the conserved BRS charge 
$Q_{\mathrm{B}}$: 
\begin{equation}
Q_{\mathrm{B}}|\mathrm{phys}\rangle =0 .
\label{eq:5}
\end{equation}
In order to clarify the particle contents of the physical states, 
we expand any field $\Phi_{A}(x)$ as 
\begin{equation}
\Phi_{A}(x) = \frac{1}{(2\pi)^{3/2}}\int\rmd^{4}p\,
\theta(p^{0})\left[\Phi_{A}(p)\rme^{\rmi px}
+\Phi_{A}^{\dag}(p)\rme^{-\rmi px}\right] .
\label{eq:6}
\end{equation} 
In the axial coordinate \ $(p^{1}=p^{2}=0,\ p^{3}>0)$, 
let us define as   
\begin{equation}
\varphi^{1}(p)\definition A^{1}(p),\ \ \ \ 
\varphi^{2}(p) \definition A^{2}(p),\ \ \ \ 
\chi(p) \definition \frac{1}{p^{3}}A^{3}(p) .
\label{eq:7}
\end{equation}
Then we find that $\left\{\varphi^{i}(p)\,(i=1,2)\right\}$ 
are BRS singlets and physical, whereas 
$\left\{\chi(p), b(p), c(p), \bar{c}(p)\right\}$ constitute 
a BRS quartet.

\subsection{Massive vector: naive model}\label{sec:2.2} 
Mass is introduced through the Proca Lagrangian
\begin{equation}
L_{m}\left[A^{\mu}\right] = 
-\frac{1}{4}F_{\mu\nu}F^{\mu\nu} 
- \frac{m^{2}}{2}A_{\mu}A^{\mu} .
\label{eq:8}
\end{equation}
This naive model of massive vector field gives 
two-point functions with the second-order massless singularities 
like  
\begin{equation}
\langle A^{\mu}A^{\nu}\rangle = 
\frac{1}{\Box - m^{2}}\left(\eta^{\mu\nu} - 
\frac{\partial^{\mu}\partial^{\nu}}{m^{2}}\right)\delta . 
\label{eq:9}
\end{equation}
The field equations
\begin{equation}
\left\{
\begin{array}{l}
(\square-m^{2})A^{\mu}=0 ,\\
\partial_{\mu}A^{\mu}=0
\label{eq:10}
\end{array}
\right.
\end{equation}
ensure that the physical degrees of freedom count three,\  
$\left\{\varphi^{i}(p)\ (i=1,2), \ \chi(p)\right\}$,\ 
in this case.

\subsection{Massive vector: Nakanishi model}\label{sec:2.3}
In order to remove the massless singularities involved 
in the naive model, 
Nakanishi \cite{NN} proposed the following type of Lagrangian:
\begin{equation}
L_{\mathrm{N}} = L_{m}\left[A^{\mu}\right]
+L^{\alpha}_{\mathrm{`GF'}}, \ \ \ \ 
L^{\alpha}_{\mathrm{`GF'}} =
b\left(\partial_{\mu}A^{\mu} + \frac{\alpha}{2}b\right) ,
\label{eq:11}
\end{equation}
where $L_{m}\left[A^{\mu}\right]$ is the Proca Lagrangian and 
$L^{\alpha}_{\mathrm{`GF'}}$ is the gauge-fixing-like one.\footnote
{The Proca Lagrangian does not show any gauge invariance. 
Therefore, adding the term $L^{\alpha}_{\mathrm{`GF'}}$ has nothing 
to do with gauge-fixing procedure.} 
For $\alpha=1$, two-point functions are then
\begin{equation}
\langle A^{\mu}A^{\nu}\rangle = 
\frac{\eta^{\mu\nu}}{\Box - m^{2}}\delta, \ \ \ \ 
\langle A^{\mu}b\rangle = 
\frac{\partial^{\mu}}{\Box - m^{2}}\delta, \ \ \ \ 
\langle bb\rangle =
-\frac{m^{2}}{\Box - m^{2}}\delta ,
\label{eq:12}
\end{equation}
which show the massless singularities have disappeared. 
Note that the Nakanishi-Lautrup field $b(x)$ is a ghost for 
$m\neq 0$. 
Physical states are picked out by the condition 
\begin{equation}
b^{(+)}(x)|\mathrm{phys}\rangle =0 ,
\label{eq:13}
\end{equation}
where $b^{(+)}(x)$ denotes the positive frequency part of $b(x)$. 
In the massive case, we can introduce the field   
\begin{equation}
\tilde{\chi}(p) \definition
\chi(p)+\rmi\frac{1}{m^{2}}b(p) ,
\label{eq:14}
\end{equation}
and find that the set of the fields 
$\left\{\varphi^{i}(p)\ (i=1,2), \ \tilde{\chi}(p)\right\}$ 
is physical and the ghost 
$b(p)$ is unphysical. 
In the massless case, on the other hand, the set of the fields 
$\left\{\varphi^{i}(p)\ (i=1,2), \ b(p)\right\}$ becomes physical 
and the field $\chi(p)$ becomes unphysical. 
Note that the field $b$ is zero-normed in this case. 
That means $b$ is not observable, 
there remaining only two observable degrees of freedom.  

\subsection{Massive vector: BRS model}\label{sec:2.4} 
The BRS model Lagrangian is constructed as follows: 
introduce an auxiliary scalar field $\theta$, 
promote the Proca Lagrangian \eref{eq:8} 
to a gauge invariant one 
by replacing $A^{\mu}$ with $A^{\mu}-\frac{1}{m}\partial^{\mu}\theta$, 
and append a certain GF+FP Lagrangian. 
The total Lagrangian is given as 
\begin{eqnarray}
L_{\mathrm{BRS}} & = 
L_{m}\left[ A^{\mu}-\frac{1}{m}\partial^{\mu}\theta\right]
+ L^{m\alpha}_{\mathrm{GF+FP}} \nonumber\\
& = 
L_{m}\left[ A^{\mu}\right]-m\theta\partial_{\mu}A^{\mu}
-\frac{1}{2}\partial_{\mu}\theta\partial^{\mu}\theta
+ L^{m\alpha}_{\mathrm{GF+FP}} .
\label{eq:15}
\end{eqnarray}
For the GF+FP Lagrangian, 
we adopt here the following massive type instead of the massless one 
\eref{eq:2}: 
\begin{eqnarray}
L^{m\alpha}_{\mathrm{GF+FP}} & = 
b\left(\partial_{\mu}A^{\mu}-m\theta
+ \frac{\alpha}{2}b\right) 
+\rmi\bar{c}\left(\square-m^{2}\right)c \nonumber\\
& = -\rmi\delta\left[\bar{c}
\left(\partial_{\mu}A^{\mu}-m\theta
+ \frac{\alpha}{2}b\right)\right] .
\label{eq:16}
\end{eqnarray}
The choice of \eref{eq:16} gives simple pole structure to 
two-point functions 
and makes it easy to investigate 
particle contents of physical states. 
The BRS transformation which keeps the theory invariant is 
\begin{equation}
\delta A_{\mu} = \partial_{\mu}c , \ \ \ \ 
\delta\theta = mc, \ \ \ \ 
\delta\bar{c} = \rmi b .
\label{eq:17}
\end{equation}
For $\alpha=1$, two-point functions are calculated to give 
\begin{equation}\fl
\left.
\begin{array}{l}
\displaystyle
\langle A^{\mu}A^{\nu}\rangle =  
\frac{\eta^{\mu\nu}}{\Box-m^{2}}\delta , \ \ \ \   
\langle A^{\mu}b\rangle = 
\frac{\partial^{\mu}}{\Box-m^{2}}\delta , \ \ \ \  
\langle bb\rangle = 0 , \ \ \ \  
\langle c\bar{c}\rangle = -\rmi\frac{1}{\Box-m^{2}}\delta , \\
\displaystyle\makebox[23mm]{}
\langle A^{\mu}\theta\rangle = 0, \ \ \ \ 
\langle b\theta\rangle = \frac{m}{\Box-m^{2}}\delta , \ \ \ \ 
\langle\theta\theta\rangle = \frac{1}{\Box-m^{2}}\delta . 
\end{array}
\right.
\label{eq:18}
\end{equation}
Physical states are defined as
\begin{equation}
Q_{\mathrm{B}}|\mathrm{phys}\rangle =0 .
\label{eq:19}
\end{equation}
If we introduce the field 
\begin{equation}
\tilde{\theta}(p)\definition \theta(p)+\rmi m\chi(p)+
\frac{m}{(p^{3})^{2}}b(p) ,
\label{eq:20}
\end{equation}
we can find the following particle contents:  
$\{\varphi^{i}(p)\ (i=1,2), \,\tilde{\theta}(p)\}$ are 
BRS singlets and physical; 
and $\{\chi(p), b(p), c(p), \bar{c}(p)\}$ make up a BRS quartet. 

\section{Massive tensor: linear theories}\label{sec:3}
\subsection{Massless tensor}\label{sec:3.1}
A massless tensor field is described by the Lagrangian
\begin{equation}
L_{0} = \frac{1}{2}h^{\mu\nu}\Lambda_{\mu\nu ,\rho\sigma}h^{\rho\sigma} 
+ L^{\alpha}_{\mathrm{GF+FP}} 
\label{eq:21}
\end{equation}
with  
\begin{equation}\fl
\Lambda_{\mu\nu ,\rho\sigma} \definition 
\left(\eta_{\mu\rho}\eta_{\nu\sigma} - \eta_{\mu\nu}
\eta_{\rho\sigma}
\right)\square 
- \left(\eta_{\mu\rho}\partial_{\nu}
\partial_{\sigma} 
+ \eta_{\nu\sigma}\partial_{\mu}\partial_{\rho}\right)
+ \left(\eta_{\rho\sigma}\partial_{\mu}\partial_{\nu} 
+ \eta_{\mu\nu}\partial_{\rho}\partial_{\sigma}\right) .
\label{eq:22}
\end{equation}
For the GF+FP Lagrangian, we choose   
\begin{eqnarray}
L^{\alpha}_{\mathrm{GF+FP}} & = 
b_{\mu}\left(
\partial_{\nu}h^{\mu\nu} - \frac{1}{2}\partial^{\mu}h 
+ \frac{\alpha}{2}b^{\mu}\right) 
+\rmi\bar{c}_{\mu}\square\,c^{\mu} \nonumber\\
& = -\rmi\delta\left[\bar{c}_{\mu}
\left(\partial_{\nu}h^{\mu\nu} - \frac{1}{2}\partial^{\mu}h 
+ \frac{\alpha}{2}b^{\mu}\right)\right] ,
\label{eq:23}
\end{eqnarray}
where $\alpha$ is the gauge parameter and 
$h$ is defined by $h\definition h^{\mu}_{\mu}$. 
The theory is invariant under the BRS transformation 
\begin{equation}
\delta h^{\mu\nu} = \partial^{\mu}c^{\nu} + 
\partial^{\nu}c^{\mu} , \ \ \ \ 
\delta\bar{c}_{\mu} = \rmi b_{\mu} .
\label{eq:24}
\end{equation}
For $\alpha=\frac{1}{2}$, we have simple forms of
 two-point functions: 
\begin{equation}
\left.
\begin{array}{l}
\displaystyle
\makebox[-3mm]{}
\langle h^{\mu\nu}h^{\rho\sigma}\rangle = 
\frac{1}{\square}\,\frac{1}{2}\left(
\eta^{\mu\rho}\eta^{\nu\sigma} 
+ \eta^{\mu\sigma}\eta^{\nu\rho} 
- \eta^{\mu\nu}\eta^{\rho\sigma}\right)\delta, \\[3mm]
\displaystyle
\makebox[-3mm]{}
\langle h^{\mu\nu}b^{\rho}\rangle = 
\frac{1}{\square}\left(\eta^{\mu\rho}\partial^{\nu}
+\eta^{\nu\rho}\partial^{\mu}\right)\delta, \ \ \ \
\langle b^{\mu}b^{\nu}\rangle = 0, \ \ \ \ 
\langle c^{\mu}\bar{c}_{\nu}\rangle =
-\rmi\frac{1}{\square}\,\delta^{\mu}_{\nu}\delta .
\end{array}
\right. 
\label{eq:25}
\end{equation}
Physical states are defined by the condition 
\begin{equation}
Q_{\mathrm{B}}|\mathrm{phys}\rangle =0 .
\label{eq:26}
\end{equation}
In the axial coordinate $(p^{1}=p^{2}=0,\ p^{3}>0)$\ 
we define 
\begin{equation}\fl
\left.
\begin{array}{l}
\displaystyle\makebox[23mm]{}
\phi^{1}(p)\definition \frac{1}{2}\left[
h^{11}(p)-h^{22}(p)\right],\ \ \ \ 
\phi^{2}(p) \definition h^{12}(p),\\[2mm]
\displaystyle
\chi^{0}(p) \definition \frac{1}{2p^{0}}h^{00}(p), \ \ \ 
\chi^{1}(p) \definition \frac{1}{p^{0}}h^{01}(p), \ \ \ 
\chi^{2}(p) \definition \frac{1}{p^{0}}h^{02}(p), \ \ \ 
\chi^{3}(p) \definition \frac{1}{2p^{3}}h^{33}(p) .
\end{array}
\right.
\label{eq:27}
\end{equation}
Particle contents are then as follows: 
$\{\phi^{i}(p)\ (i=1,2)\}$ are BRS singlets and physical; 
$\{\chi^{\mu}(p), b_{\mu}(p), c^{\mu}(p), \bar{c}_{\mu}(p)\}$\ 
form BRS quartets.

\subsection{Massive tensor: naive model}\label{sec:3.2}
Naive introduction of mass is carried out through the Lagrangian
\begin{equation}
L_{m}^{a}\left[ h^{\mu\nu}\right] = 
\frac{1}{2}h^{\mu\nu}\Lambda_{\mu\nu ,\rho\sigma}h^{\rho\sigma} 
- \frac{m^{2}}{2}\left( h^{\mu\nu}h_{\mu\nu} - ah^{2}\right) .
\label{eq:28}
\end{equation}
The parameter $a$ has two choices of interest:\ 
$a=1$ and $a=\frac{1}{2}$, 
corresponding to the PT model and 
the ASG one respectively. 
In the case of $a=1$, 
the Lagrangian has the Fierz-Pauli type mass term,  
and gives the field equations 
\begin{equation}
\left\{
\begin{array}{l}
\left(\square-m^{2}\right)h^{\mu\nu}=0 , \\
\partial_{\nu}h^{\mu\nu}=0  , \\
h=0 .
\end{array} 
\right.
\label{eq:29}
\end{equation}
Therefore, this model does describe an ordinary massive tensor field  
of five degrees of freedom. 
The two-point functions 
\begin{eqnarray}\fl
\langle h^{\mu\nu}h^{\rho\sigma}\rangle = &  
\!\!\!\!\!\frac{1}{\square - m^{2}}\left\{
\frac{1}{2}\left(
\eta^{\mu\rho}\eta^{\nu\sigma} 
+ \eta^{\mu\sigma}\eta^{\nu\rho} 
- \eta^{\mu\nu}\eta^{\rho\sigma}\right) \right. \nonumber\\
& \makebox[15mm]{}
- \frac{1}{2m^{2}}\left(
\eta^{\mu\rho}\partial^{\nu}\partial^{\sigma}
+ \eta^{\mu\sigma}\partial^{\nu}\partial^{\rho}
+ \eta^{\nu\rho}\partial^{\mu}\partial^{\sigma}
+ \eta^{\nu\sigma}\partial^{\mu}\partial^{\rho}\right) 
\nonumber\\
& \makebox[14mm]{}\left.\mbox{}
+ \frac{2}{3}\left(
\frac{1}{2}\eta^{\mu\nu} 
+ \frac{\partial^{\mu}\partial^{\nu}}{m^{2}}\right)\left(
\frac{1}{2}\eta^{\rho\sigma} 
+ \frac{\partial^{\rho}\partial^{\sigma}}{m^{2}}\right)
\right\}\delta
\label{eq:30}
\end{eqnarray}
show the fourth-order massless singularities. 
In the case of $a=\frac{1}{2}$, on the other hand, 
field equations reduce to 
\begin{equation}
\left\{
\begin{array}{l}
\left(\square-m^{2}\right)h^{\mu\nu}=0 , \\
\displaystyle
\partial_{\nu}h^{\mu\nu}-\frac{1}{2}\partial^{\mu}h=0 .
\end{array} 
\right.
\label{eq:31}
\end{equation}
The number of physical degrees of freedom of this model is six; 
five corresoponds to a massive tensor, and
one is to an additional scalar ghost field. 
The two-point functions  
\begin{eqnarray}\fl
\langle h^{\mu\nu}h^{\rho\sigma}\rangle = 
\frac{1}{\square - m^{2}}\left\{
\frac{1}{2}\left(
\eta^{\mu\rho}\eta^{\nu\sigma} 
+ \eta^{\mu\sigma}\eta^{\nu\rho} 
- \eta^{\mu\nu}\eta^{\rho\sigma}\right) \right. \nonumber\\
\makebox[14mm]{}\left.\mbox{}
- \frac{1}{2m^{2}}\left(
\eta^{\mu\rho}\partial^{\nu}\partial^{\sigma}
+ \eta^{\mu\sigma}\partial^{\nu}\partial^{\rho}
+ \eta^{\nu\rho}\partial^{\mu}\partial^{\sigma}
+ \eta^{\nu\sigma}\partial^{\mu}\partial^{\rho}\right)
\right\}\delta 
\label{eq:32}
\end{eqnarray}
contain only the second-order massless singularities in this case. 

\subsection{ASG-type massive tensor: BRS model}\label{sec:3.3}
In order to promote the ASG model to a BRS invariant one, 
introduce an auxiliary vector field $\theta^{\mu}$, 
replace $h^{\mu\nu}$ with the combination\ 
$h^{\mu\nu}-\frac{1}{m}(\partial^{\mu}\theta^{\nu}
-\partial^{\nu}\theta^{\mu})$ 
in the Lagrangian \eref{eq:28} with $a=\frac{1}{2}$, 
and append a certain GF+FP Lagrangian. 
The total Lagrangian is  
\begin{eqnarray}
L_{\mathrm{BRS}}^{a=\frac{1}{2}} & = 
L_{m}^{a=\frac{1}{2}}\left[
h^{\mu\nu} - \frac{1}{m}\left(
\partial^{\mu}\theta^{\nu} + \partial^{\nu}\theta^{\mu}\right)
\right] + L^{m\alpha}_{\mathrm{GF+FP}} 
\nonumber\\
& = L_{m}^{a=\frac{1}{2}}\left[ h^{\mu\nu}\right]
- 2m\theta_{\mu}\left(
\partial_{\nu}h^{\mu\nu} - \frac{1}{2}\partial^{\mu}h\right) 
- \partial_{\mu}\theta_{\nu}\partial^{\mu}\theta^{\nu} 
+ L^{m\alpha}_{\mathrm{GF+FP}} . 
\label{eq:33}
\end{eqnarray}
Following the BRS procedure for the massive vector 
in \sref{sec:2.4}, 
we adopt the following GF+FP Lagrangian of massive type: 
\begin{eqnarray}
L^{m\alpha}_{\mathrm{GF+FP}} & =
b_{\mu}\left(\partial_{\nu}h^{\mu\nu}
-\frac{1}{2}\partial^{\mu}h-m\theta^{\mu}
+\frac{\alpha}{2}b^{\mu}\right)
+\rmi\bar{c}_{\mu}\left(\square -m^{2}\right)c^{\mu}
\nonumber\\
& = -\rmi\delta\left[\bar{c}_{\mu}\left(
\partial_{\nu}h^{\mu\nu}
-\frac{1}{2}\partial^{\mu}h-m\theta^{\mu}
+\frac{\alpha}{2}b^{\mu}\right)\right] .
\label{eq:34}
\end{eqnarray}
The BRS transformation  
\begin{equation}
\delta h^{\mu\nu} = \partial^{\mu}c^{\nu} + 
\partial^{\nu}c^{\mu} , \ \ \ 
\delta\theta^{\mu} = mc^{\mu} , \ \ \ 
\delta\bar{c}_{\mu} = \rmi b_{\mu} 
\label{eq:35}
\end{equation}
keeps the system invariant. 
For $\alpha=\frac{1}{2}$, 
two-point functions are calculated as 
\begin{equation}\fl
\left.
\begin{array}{l}
\displaystyle\makebox[23mm]{}
\langle h^{\mu\nu}h^{\rho\sigma}\rangle = 
\frac{1}{\square -m^{2}}\,\frac{1}{2}\left(
\eta^{\mu\rho}\eta^{\nu\sigma} 
+ \eta^{\mu\sigma}\eta^{\nu\rho} 
- \eta^{\mu\nu}\eta^{\rho\sigma}\right)\delta, \\[3mm]
\displaystyle
\langle h^{\mu\nu}b^{\rho}\rangle = 
\frac{1}{\square -m^{2}}\left(\eta^{\mu\rho}\partial^{\nu}
+\eta^{\nu\rho}\partial^{\mu}\right)\delta, \ \ \ \
\langle b^{\mu}b^{\nu}\rangle = 0, \ \ \ \ 
\langle c^{\mu}\bar{c}_{\nu}\rangle =
-\rmi\frac{1}{\square -m^{2}}\,\delta^{\mu}_{\nu}\delta , \\[3mm]
\displaystyle\makebox[23mm]{}
\langle h^{\mu\nu}\theta^{\rho}\rangle = 0, \ \ \ \ 
\langle b^{\mu}\theta^{\nu}\rangle =
\frac{m}{\square -m^{2}}\eta^{\mu\nu}\delta, \ \ \ \ 
\langle\theta^{\mu}\theta^{\nu}\rangle =
\frac{1}{2}\,\frac{1}{\square -m^{2}}\eta^{\mu\nu}\delta .
\end{array}
\right.
\label{eq:36}
\end{equation}
Physical states are defined by
\begin{equation}
Q_{\mathrm{B}}|\mathrm{phys}\rangle =0 .
\label{eq:37}
\end{equation}
Let us introduce a field $\tilde{\theta}^{\mu}(p)$ as 
the combination 
\begin{equation}
\tilde{\theta}^{\mu}(p)\definition
\theta^{\mu}(p)+\rmi m\chi^{\mu}(p)+m\omega^{\mu\nu}b_{\nu}(p) 
\label{eq:38}
\end{equation}
with the matrix $\omega^{\mu\nu}$ having the components   
\begin{equation}
\left\{
\begin{array}{l}
\displaystyle
\omega^{00}=\frac{1}{8(p^{0})^{2}}, \ \ 
\omega^{03}=\omega^{30}=\frac{1}{8p^{0}p^{3}}, \ \  
\omega^{33}=\frac{1}{8(p^{3})^{2}}, \\
\displaystyle 
\omega^{11}=\omega^{22}=-\frac{1}{2(p^{0})^{2}}, \ \  
\mathrm{the\ others}=0 .
\end{array}
\right.
\label{eq:39}
\end{equation}
Then we find the following particle contents: 
$\{\phi^{i}(p)\ (i=1,2),\,\tilde{\theta}^{\mu}(p)\}$\ 
are BRS singlets and physical;  
$\{\chi^{\mu}(p), b_{\mu}(p), c^{\mu}(p), \bar{c}_{\mu}(p)\}$\ 
make up BRS quartets. 
Note that there remains a ghost $\tilde{\theta}^{0}(p)$\  
in the physical states. 
Nonlinear completion of this model 
including a possible mechanism of killing the ghost 
was reported at QTS-4 
\cite{SH6} (See also \cite{SH5}.).

\subsection{PT-type massive tensor: 
BRS+Nakanishi model}\label{sec:3.4}
We have seen in \sref{sec:3.2} that the PT model of massive tensor shows 
fourth-order massless singularities in two-point functions. 
Those singularities cannot be removed by such simple application 
of the BRS procedure as done in \sref{sec:3.3}. 
So we invoke the Nakanishi procedure in addition to the BRS one.  
For the Lagrangian, we adopt the following form: 
\begin{eqnarray}
L_{\mathrm{BRS+N}}^{a=1} & = 
L_{m}^{a=1}\left[ h^{\mu\nu} - \frac{1}{m}\left(
\partial^{\mu}\theta^{\nu} + \partial^{\nu}\theta^{\mu}
\right)\right]
+ L^{m\alpha}_{\mathrm{GF+FP}} 
+ L^{\beta}_{\mathrm{`GF'}}\nonumber \\
& = L_{m}^{a=1}\left[ h^{\mu\nu}\right]
- 2m\theta_{\mu}
\left(\partial_{\nu}h^{\mu\nu} - \partial^{\mu}h\right) 
- \frac{1}{2}\left(\partial^{\mu}\theta^{\nu} - \partial^{\nu}
\theta^{\mu}\right)^{2} \nonumber \\
& \makebox[60mm]{}
+ L^{m\alpha}_{\mathrm{GF+FP}} 
+L^{\beta}_{\mathrm{`GF'}} .
\label{eq:40}
\end{eqnarray}
Here the first and the second terms 
on the right side of the first line are from the 
BRS procedure, and the third term $L^{\beta}_{\mathrm{`GF'}}$\ 
represents the gauge-fixing-like term in the Nakanishi procedure. 
For $L^{\beta}_{\mathrm{`GF'}}$, we also choose 
massive type of the following form:   
\begin{eqnarray}
L^{\beta}_{\mathrm{`GF'}} & =
b\left(\partial_{\mu}\theta^{\mu}
-\frac{m}{2}h+\frac{\beta}{2}b\right)
\label{eq:41}
\end{eqnarray}
with the second parameter $\beta$. 
Assuming the Nakanishi-Lautrup field $b$ is BRS invariant, 
the total Lagrangian is invariant under the BRS transformation  
\begin{equation}
\delta h^{\mu\nu} = \partial^{\mu}c^{\nu} + 
\partial^{\nu}c^{\mu} , \ \ \  
\delta\theta^{\mu} = mc^{\mu} , \ \ \  
\delta\bar{c}_{\mu} = \rmi b_{\mu}, \ \ \  
\delta b = 0 .
\label{eq:42}
\end{equation}
Two-point functions show simple pole structure for 
$\alpha=\beta=\frac{1}{2}$ as follows:
\begin{equation}\fl
\left.
\begin{array}{l}
\displaystyle\makebox[23mm]{}
\langle h^{\mu\nu}h^{\rho\sigma}\rangle = 
\frac{1}{\square -m^{2}}\,\frac{1}{2}\left(
\eta^{\mu\rho}\eta^{\nu\sigma} 
+ \eta^{\mu\sigma}\eta^{\nu\rho} 
- \eta^{\mu\nu}\eta^{\rho\sigma}\right)\delta, \\[3mm]
\displaystyle
\langle h^{\mu\nu}b^{\rho}\rangle = 
\frac{1}{\square -m^{2}}\left(\eta^{\mu\rho}\partial^{\nu}
+\eta^{\nu\rho}\partial^{\mu}\right)\delta, \ \ \ \
\langle b^{\mu}b^{\nu}\rangle = 0, \ \ \ \ 
\langle c^{\mu}\bar{c}_{\nu}\rangle =
-\rmi\frac{1}{\square -m^{2}}\,\delta^{\mu}_{\nu}\delta ,\\[3mm]
\displaystyle\makebox[23mm]{}
\langle h^{\mu\nu}\theta^{\rho}\rangle = 0, \ \ \ \ 
\langle b^{\mu}\theta^{\nu}\rangle =
\frac{m}{\square -m^{2}}\eta^{\mu\nu}\delta, \ \ \ \ 
\langle\theta^{\mu}\theta^{\nu}\rangle =
\frac{1}{2}\,\frac{1}{\square -m^{2}}\eta^{\mu\nu}\delta ,\\[3mm]
\displaystyle
\langle h^{\mu\nu}b\rangle = 
-\frac{m}{\square -m^{2}}\eta^{\mu\nu}\delta , \ \ \ \ 
\langle\theta^{\mu}b\rangle = 
\frac{\partial^{\mu}}{\square -m^{2}}\delta , \ \ \ \ 
\langle b^{\mu}b\rangle = 0 , \ \ \ \ 
\langle bb\rangle = -\frac{6m^{2}}{\square -m^{2}}\delta .
\end{array}
\right.
\label{eq:43}
\end{equation}
The fourth-order massless singularities have been driven away indeed. 
Note that, as seen from the last equation of \eref{eq:43}, 
the field $b$ is a ghost for $m\neq 0$.  
This is the same situation as in the case of massive vector 
in \sref{sec:2.3}.
Physical states are picked out by two conditions of the BRS type and 
the Nakanishi type: 
\begin{equation}
Q_{\mathrm{B}}|\mathrm{phys}\rangle =0 , \ \ \ \ 
b^{(+)}(x)|\mathrm{phys}\rangle =0 .
\label{eq:44}
\end{equation}
In order to investigate the particle contents, we introduce 
the following quantities: 
\begin{equation}
\varphi^{1}(p)\definition\tilde{\theta}^{1}(p), \ \ \ \ 
\varphi^{2}(p)\definition\tilde{\theta}^{2}(p), \ \ \ \ 
\chi (p)\definition\frac{1}{p^{3}}\tilde{\theta}^{3}(p) .
\label{eq:45}
\end{equation}
For the massive case, we can introduce the combination 
\begin{equation}
\tilde{\chi}(p) \definition
\chi(p)+\rmi\,\frac{1}{6}\left(
\frac{1}{m^{2}}+\frac{1}{2(p^{3})^{2}}\right)b(p) .
\label{eq:46}
\end{equation}
Particle contents are then:  
$\{\phi^{i}(p), \ \varphi^{i}(p)\ (i=1,2), \ \tilde{\chi}(p)\}$\ 
are BRS singlets and physical; 
$b(p)$ is a BRS singlet but unphysical (\textit{ghost});\ 
$\{\chi^{\mu}(p), b_{\mu}(p), c^{\mu}(p), \bar{c}_{\mu}(p)\}$\ 
constitute BRS quartets. 
For the massless case, on the other hand, 
we cannot define a field like $\tilde{\chi}$. 
In this case, we find the following particle contents: 
$\{\phi^{i}(p), \ \varphi^{i}(p)\ (i=1,2), \ b(p)\}$\ 
are BRS singlets and physical; 
$\chi(p)$ is a BRS singlet but unphysical;\ 
$\{\chi^{\mu}(p), b_{\mu}(p), c^{\mu}(p), \bar{c}_{\mu}(p)\}$\ 
form BRS\ quartets.
Note again that in the massless case, 
$b$ is physical but unobservable because it is zero-normed. 
From now on, we focus on the model 
described by the Lagrangian \eref{eq:40} with 
$\alpha=\beta=\frac{1}{2}$. 

\section{Massive tensor: nonlinear completion}\label{sec:4}
\subsection{Nonlinear BRS transformation}\label{sec:4.1}
To study nonlinear theories we introduce 
the metric $g_{\mu\nu}$ and the tetrad $e_{k}^{\ \mu}$ through  
\begin{equation}
g_{\mu\nu} \definition \eta_{\mu\nu}-\kappa h_{\mu\nu}, 
\;\;\;\;\;e_{k}^{\ \mu}e^{k\nu}=g^{\mu\nu} 
\label{eq:47} 
\end{equation}
with the gravitational constant $\kappa$. 
The linear BRS transformation \eref{eq:42} is extended to its 
nonlinear form:  
\begin{equation}
\left\{
\begin{array}{l}
\displaystyle
\delta e_{k}^{\ \mu} = \kappa\left(
\partial_{\rho}c^{\mu}\cdot e_{k}^{\ \rho}
-c^{\rho}\partial_{\rho}e_{k}^{\ \mu}\right) ,\\ 
\displaystyle
\delta\theta^{\mu} = mc^{\mu}
- \kappa c^{\rho}\partial_{\rho}\theta^{\mu} ,\\
\displaystyle
\delta c^{\mu} = -\kappa c^{\rho}\partial_{\rho}c^{\mu} ,\\
\displaystyle
\delta\bar{c}_{\mu} = \rmi b_{\mu} ,\\
\displaystyle
\delta b = -\kappa c^{\rho}\partial_{\rho}b .
\end{array}
\right.
\label{eq:48}
\end{equation}
Basic quantities invariant under the nonlinear BRS transformation
can be constructed as 
\begin{equation}
E_{k}^{\ \mu} \definition 
e_{k}^{\ \mu}-
\frac{\kappa}{m}e_{k}^{\ \rho}\partial_{\rho}\theta^{\mu} ,
\label{eq:49}
\end{equation}
\begin{equation}
G^{\mu\nu} \definition 
E_{k}^{\ \mu}E^{k\nu}
= g^{\mu\nu}-\frac{\kappa}{m}\left(
g^{\rho\mu}\partial_{\rho}\theta^{\nu}
+g^{\rho\nu}\partial_{\rho}\theta^{\mu}\right)
+\left(\frac{\kappa}{m}\right)^{2}
g^{\rho\sigma}\partial_{\rho}\theta^{\mu}
\partial_{\sigma}\theta^{\nu} .
\label{eq:50}
\end{equation}
In fact they behave as scalars under the transformation \eref{eq:48}: 
\begin{equation}
\delta E_{k}^{\ \mu}= 
-\kappa c^{\rho}\partial_{\rho}E_{k}^{\ \mu}, \ \ \ \ 
\delta G^{\mu\nu} = -\kappa c^{\rho}\partial_{\rho}G^{\mu\nu} .
\label{eq:51} 
\end{equation}
Possible Lagrangians are therefore of the form 
\begin{equation}
L = \sqrt{-g}\,F\left( E_{k}^{\ \mu}, b\right) ,
\label{eq:52}
\end{equation}
where $F$ is an arbitrary function. 
The action is indeed invariant, 
because such Lagrangian as \eref{eq:52} is transformed as 
\begin{equation}
\delta L = -\kappa\partial_{\mu}\left(c^{\mu}L\right) .
\label{eq:53}
\end{equation}

\subsection{Nonlinear Lagrangian}\label{sec:4.2} 
We require for the Lagrangian 
to be at most quadratic in $E_{k}^{\ \mu}$  
\textit{and} to reduce to $L_{\mathrm{BRS+N}}^{a=1}$\ 
in the flat-spacetime limit $(\kappa\rightarrow 0)$.  
Then we have the following form consisting of four terms: 
\begin{equation}
L = \tilde{L}_{m}+\gamma\tilde{L}_{\mathrm{R}}
+\tilde{L}^{\alpha}_{\mathrm{GF+FP}}
+\tilde{L}^{\beta}_{\mathrm{`GF'}} ,
\label{eq:54}
\end{equation}
with an arbitrary real number $\gamma$. 
These terms are given by 
\begin{equation}
\makebox[-5mm]{}
\tilde{L}_{m} = \frac{1}{2\kappa^{2}}\sqrt{-g}\left\{ 
R+\frac{m^{2}}{2}\left[ 6-G^{\mu\nu}\eta_{\mu\nu}
-\left( E_{k}^{\ \mu}\delta_{\mu}^{k}\right)^{2}
+2E_{k}^{\ \mu}\delta_{\mu}^{l}E_{l}^{\ \nu}\delta_{\nu}^{k}
\right]\right\} ,
\label{eq:55}
\end{equation}
\begin{equation}
\makebox[-5mm]{}
\tilde{L}_{\mathrm{R}} = 
\frac{m^{2}}{\kappa^{2}}\sqrt{-g}\left[
-3+2E_{k}^{\ \mu}\delta_{\mu}^{k}
-\frac{1}{2}\left(E_{k}^{\ \mu}\delta_{\mu}^{k}\right)^{2}
+\frac{1}{2}E_{k}^{\ \mu}\delta_{\mu}^{l}E_{l}^{\ \nu}\delta_{\nu}^{k}
\right] ,
\label{eq:56}
\end{equation}
\begin{eqnarray}
\makebox[-5mm]{}
\tilde{L}^{\alpha}_{\mathrm{GF+FP}} & = 
-\rmi\delta\left[\bar{c}_{\mu}\left(
\frac{1}{\kappa}\partial_{\nu}\tilde{g}^{\mu\nu}
-m\theta^{\mu}
+\frac{\alpha}{2}\eta^{\mu\nu}b_{\nu}\right)\right]
\nonumber \\
& = 
b_{\mu}\left(\frac{1}{\kappa}\partial_{\nu}\tilde{g}^{\mu\nu}
-m\theta^{\mu}
+\frac{\alpha}{2}\eta^{\mu\nu}b_{\nu}\right) 
+ \rmi\bar{c}_{\mu}\left(
\partial_{\nu}D^{\mu\nu}_{\ \ \,\rho}-m^{2}\delta^{\mu}_{\rho}
\right)c^{\rho}
\nonumber \\
& \makebox[60mm]{}
+\rmi\kappa m\bar{c}_{\mu}c^{\rho}\partial_{\rho}\theta^{\mu} ,
\label{eq:57}
\end{eqnarray}
and 
\begin{equation}
\tilde{L}^{\beta}_{\mathrm{`GF'}}
=\sqrt{-g}\,b\left[
\frac{m}{\kappa}\left(\delta^{\mu}_{k}-E_{k}^{\ \mu}\right)
\delta^{k}_{\mu}+\frac{\beta}{2}b\,\right] ,
\label{eq:58}
\end{equation}
where we have used the definitions 
\begin{eqnarray}
\tilde{g}^{\mu\nu} &\definition\sqrt{-g}\,g^{\mu\nu} ,
\label{eq:59}
\\
D^{\mu\nu}_{\ \ \,\rho} &\definition 
\tilde{g}^{\mu\sigma}\delta_{\rho}^{\nu}\partial_{\sigma}
+\tilde{g}^{\nu\sigma}\delta_{\rho}^{\mu}\partial_{\sigma}
-\tilde{g}^{\mu\nu}\partial_{\rho}
-\left(\partial_{\rho}\tilde{g}^{\mu\nu}\right) .
\label{eq:60}
\end{eqnarray}
We can easily verify that the main part of the Lagrangian 
$\tilde{L}_{m}+\tilde{L}^{\alpha}_{\mathrm{GF+FP}} 
+\tilde{L}^{\beta}_{\mathrm{`GF'}}$ goes to 
$L_{\mathrm{BRS+N}}^{a=1}$ and the redundant part 
$\tilde{L}_{\mathrm{R}}$ becomes null 
in the flat spacetime limit, 
$\kappa\rightarrow 0$. 

\subsection{Physical states}\label{sec:4.3}
In the linear theory, physical states are picked out 
by the two conditions, 
the BRS-type one $Q_{\mathrm{B}}|\mathrm{phys}\rangle =0$ and 
the Nakanishi-type one $b^{(+)}(x)|\mathrm{phys}\rangle =0$, 
as stated in \sref{sec:3.4}. 
Going to the nonlinear theory, 
the BRS-type condition takes over the same form:  
\begin{equation}
Q_{\mathrm{B}}|\mathrm{phys}\rangle =0 .
\label{eq:61}
\end{equation}
However, it is not an easy task to find a nonlinear version 
of the Nakanishi-type condition: 
\begin{equation}
\textrm{``}b^{(+)}(x)\textrm{''}|\mathrm{phys}\rangle =0 .
\label{eq:62}
\end{equation}
The problem is how to define ``$b^{(+)}(x)$'' in the nonlinear case.
In the linear case, $b(x)$ satisfies the free field equation 
$\left(\square -m^{2}\right)b(x)=0$.
This fact allows to impose the physical-state  
condition of the Nakanishi type 
$b^{(+)}(x)|\mathrm{phys}\rangle =0$.
In the nonlinear case, however, $b(x)$ obeys some 
nonlinear equation.
Setting up an auxiliary condition consistently in that 
case is still unsolved.

\section{Summary and discussion}\label{sec:5}
We have presented a possible nonlinear 
completion of massive gravity of the Fierz-Pauli type. 
Physical implications of this model are under study.

This model has the smooth massless $(m\rightarrow 0)$\ 
as well as the smooth flat-spacetime $(\kappa\rightarrow 0)$\ 
limits. 
In the flat-spacetime limit, it reduces to the 
BRS+Nakanishi extension of the PT (Fierz-Pauli) model. 

Finding a nonlinear version of the Nakanishi-type 
physical-state condition is still unsettled.

\end{document}